\documentclass[twocolumn,english]{revtex4-1}
\usepackage[T1]{fontenc}
\usepackage[latin9]{inputenc}
\usepackage{textcomp}
\usepackage{amsmath}
\usepackage{amssymb}
\usepackage{esint}

\makeatletter
\usepackage[colorlinks = true,
            linkcolor = red,
            urlcolor  = blue,
            citecolor = blue,
            anchorcolor = blue]{hyperref}

\makeatother

\usepackage{babel}
\begin{document}

\title{\texttt{TWO-DIMENSIONAL PAULI EQUATION IN NONCOMMUTATIVE PHASE-SPACE}}

\author{{\normalsize{}Ilyas Haouam}}

\email{ilyashaouam@live.fr ; ilyashaouam@ymail.com}

\address{Laboratoire de Physique Mathématique et de Physique Subatomique (LPMPS),
Université Frères Mentouri, Constantine 25000, Algeria}
\begin{abstract}
{\normalsize{}In this paper, we investigated the Pauli equation in
a two-dimensional noncommutative phase-space by considering a constant
magnetic field perpendicular to the plane. We mapped the noncommutative
problem to the equivalent commutative one through a set of two-dimensional
Bopp-shift transformation. The energy spectrum and the wave function
of the two-dimensional noncommutative Pauli equation are found, where
the problem in question has been mapped to the Landau problem. Further,
within the classical limit, we have derived the noncommutative semi-classical
partition function of the two-dimensional Pauli system of one-particle
and N-particle systems. Consequently, we have studied its thermodynamic
properties, i.e. the Helmholtz free energy, mean energy, specific
heat and entropy in noncommutative and commutative phase-spaces. The
impact of the phase-space noncommutativity on the Pauli system is
successfully examined. }{\normalsize \par}

{\normalsize{}$\phantom{}$}{\normalsize \par}

{\normalsize{}$\phantom{}$}{\normalsize \par}

\textbf{\normalsize{}Keywords:}{\normalsize{} Noncommutative phase-space,
Pauli equation, Bopp-shift, semi-classical function partition, thermodynamic
properties.}{\normalsize \par}
\end{abstract}


\maketitle

\section{{\normalsize{}Introduction}}

In the last few years, there has been a growing interest in the study
of two-dimensional systems, which have become an active area of research
because of its implications in nanofabrication technology. Such as
in graphene \cite{key-1,key-2} and other materials like Weyl semimetals
\cite{key-3}, semiconductor quantum wells, quantum Hall and fractional
Hall effects \cite{key-4,key-5}, as well the Dirac relativistic oscillator
\cite{key-6}, etc. However, despite their experimental success, it
is very important to understand these systems from a theoretical point
of view in which quantum mechanics plays a central role. Motivated
by the efforts to understand string theory \cite{key-7}, black hole
models and describe the quantum gravitation \cite{key-8,key-9,key-10}
using noncommutative geometry and by trying to have drawn considerable
attention to the phenomenological implications, we concentrate on
studying the problem of a non-relativistic spin-1/2 particle in the
presence of an electromagnetic field within two-dimensional noncommutative
phase-space. Besides, there are a large amount of articles has been
devoted to the study physics within noncommutative geometry, particularly
in quantum field theory \cite{key-11,key-12} and quantum mechanics
\cite{key-13,key-14}. 

We present the essential formulas of noncommutative algebra we need
in this manuscript. At very tiny scales (the string scale), the position
coordinates do not commute with each other, neither do the momenta.

In the two-dimensional noncommutative phase-space, the operators of
coordinates $x_{j}^{nc}$ and momenta $p_{j}^{nc}$ satisfy the following
Heisenberg-like commutation relations
\begin{equation}
\begin{array}{cccccc}
\left[x_{j}^{nc},x_{k}^{nc}\right] & = & \left[x_{j},x_{k}\right]_{\star} & = & i\Theta\epsilon_{jk}\\
\left[p_{j}^{nc},p_{k}^{nc}\right] & = & \left[p_{j},p_{k}\right]_{\star} & = & i\eta\epsilon_{jk} & ,\:(j,k=1,2)\\
\left[x_{j}^{nc},p_{k}^{nc}\right] & = & \left[x_{j},p_{k}\right]_{\star} & = & i\tilde{\hbar}\delta_{jk}
\end{array}.\label{eq:1-2}
\end{equation}

The noncommutative phase-space can be obtained using ordinary coordinates
$x_{j}$ and momenta $p_{j}$ operators and with replacing the ordinary
product by the Moyal $\star$product, which can be used as follows
\cite{key-15} 
\begin{equation}
\begin{array}{c}
\mathcal{F}\left(x^{nc},p^{nc}\right)\mathcal{G}\left(x^{nc},p^{nc}\right)=\mathcal{F}\left(x,p\right)\star\mathcal{G}\left(x,p\right)\\
=e^{\frac{i}{2}\left[\Theta_{ab}\partial_{x_{a}}\partial_{x_{b}}+\eta_{ab}\partial_{p_{a}}\partial_{p_{b}}\right]}\mathcal{F}\left(x_{a},p_{a}\right)\mathcal{G}\left(x_{b},p_{b}\right),
\end{array}\label{eq:1-1-1}
\end{equation}
where $\mathcal{F}$, $\mathcal{G}$ are two functions vary in terms
of $x$, $p$ and assumed to be infinitely differentiable. The effective
Planck constant (deformed Planck constant) is given by \cite{key-16,key-17}
\begin{equation}
\tilde{\hbar}=\hbar\left(1+\frac{\Theta\eta}{4\hbar^{2}}\right),\label{eq:2}
\end{equation}
where $\frac{\Theta\eta}{4\hbar^{2}}\ll1$ is the condition of consistency
in the usual commutative spacetime quantum mechanics, it is expected
to be generally satisfied since the small parameters $\Theta$ and
$\eta$ are of second order. $\delta_{ij}$ is the identity matrix,
$\epsilon_{jk}$ is the Levi-Civita symbol, with $\epsilon_{12}=-\epsilon_{21}=1$,
$\epsilon_{11}=\epsilon_{22}=0$. And $\Theta$, $\eta$ are the real-valued
noncommutative parameters with the dimension of $length{}^{2}$, $momentum{}^{2}$
respectively, which are assumed to be extremely small. Note that experimental
and theoretical investigations on noncommutative systems of the noncommutativity
constants led to obtaining the following upper bound on the value
of the noncommutative parameters \cite{key-17}
\begin{equation}
\Theta\preceq4.10^{-40}m^{2};\;\eta\preceq1,76.10^{-61}Kg^{2}m^{2}s^{-2}.\label{eq:3-1}
\end{equation}

Besides, recent studies \cite{key-18,key-19,key-20} revealed that
the noncommutative parameters associated with different particles
are not the same in noncommutative quantum mechanics. 

The set of operators $x_{i}^{nc}$, $p_{j}^{nc}$ are related to the
set $x_{i}$, $p_{j}$ in usual quantum mechanics by a non-canonical
linear transformation referred to as Bopp-shift as follows \cite{key-21}
\begin{equation}
\begin{array}{cc}
\begin{array}{ccc}
x^{nc} & = & x-\frac{1}{2\hbar}\Theta p_{y}\\
y^{nc} & = & y+\frac{1}{2\hbar}\Theta p_{x}
\end{array};\; & \begin{array}{ccc}
p_{x}^{nc} & = & p_{x}+\frac{1}{2\hbar}\eta y\\
p_{y}^{nc} & = & p_{y}-\frac{1}{2\hbar}\eta x
\end{array}\end{array}.\label{eq:4}
\end{equation}

The quantum mechanical system will become merely the noncommutative
one using equation (\ref{eq:4}) or (\ref{eq:1-1-1}). Let $H\left(x,p\right)$
be the Hamiltonian operator of the usual quantum system, then the
corresponding noncommutative Schrödinger equation is given by
\begin{equation}
H\left(x,p\right)\star\psi\left(x,p\right)=H\left(x_{i}-\frac{\Theta_{ij}}{2\hbar}p_{j},\;p_{i}+\frac{\eta_{ij}}{2\hbar}x_{j}\right)\psi=E\psi.\label{eq:5}
\end{equation}

Noting that noncommutative term always can be treated as a perturbation
in quantum mechanics.

In the ordinary two-dimensional commutative phase-space, the canonical
variables $x_{j}$ and $p_{i}$ satisfy the following canonical commutation
\begin{equation}
\begin{array}{cccc}
\left[x_{j},x_{k}\right] & = & 0\\
\left[p_{j},p_{k}\right] & = & 0 & ,\:(j,k=1,2)\\
\left[x_{j},p_{k}\right] & = & i\hbar\delta_{jk}
\end{array}.\label{eq:6}
\end{equation}

The paper is organized as follows. The formulation of the two-dimensional
noncommutative geometry is briefly outlined in section I. The exact
solution to the two-dimensional noncommutative Pauli equation is presented
in section II. Section III is devoted to present the thermodynamic
properties of the problem in question. Therefore, concluding with
some remarks.

\section{{\normalsize{}Two-dimensional noncommutative Pauli equation }}

The time-independent Pauli equation is given by \cite{key-22}
\begin{equation}
\frac{1}{2m_{e}}\left(\mathbf{p}-\frac{e}{c}\mathbf{A}\right)^{2}\psi+e\phi\psi+\mu_{B}\mathbf{\sigma B}\psi=E\psi,\label{eq:7}
\end{equation}
where $\psi=\left(\begin{array}{cc}
\psi_{1} & \psi_{2}\end{array}\right)^{T}$ is a two-component spinor, $\mathbf{p}=i\hbar\mathbf{\nabla}$ is
the momentum operator, $m_{e}$, $e$ are the mass and charge of the
electron, and $c$ is the speed of light. As well $\mu_{B}=\frac{\left|e\right|\hbar}{2mc}$
is the Bohr magneton, $\mathbf{B}$ is the applied magnetic field
vector, with $\mathbf{A}\left(\mathbf{r},t\right)$ is the vector
potential and $\phi\left(\mathbf{r},t\right)$ is the Coulomb potential.
$\mathbf{\sigma}=\left(\sigma_{x},\sigma_{y},\sigma_{z}\right)$ are
the Pauli matrices. 

The time-independent Pauli equation in noncommutative phase-space
is given by
\begin{equation}
\left\{ \frac{1}{2m_{e}}\left(\mathbf{p}^{nc}-\frac{e}{c}\mathbf{A}^{\star}\right)^{2}+e\phi+\mu_{B}\mathbf{\sigma.B}\right\} \bar{\psi}=E\bar{\psi},\label{eq:8}
\end{equation}
with $\bar{\psi}$ is the noncommutative spinor wave function. If
the magnetic field $\mathbf{B}$ oriented along the axis (Oz), which
is often referred to as the Landau system, and based on the proposal
that noncommutative observables corresponding to the commutative one
\cite{key-23}, we have the following deduced noncommutative symmetric
gauge
\begin{equation}
\mathbf{A}^{\star}=\left(A_{x}^{\star},A_{y}^{\star},A_{z}^{\star}\right)=\frac{B}{2}\left(-y^{nc},x^{nc},0\right),\:A_{0}^{\star}=e\phi^{\star}=0.\label{eq:9}
\end{equation}

Here the electron is unbound $\phi=0$. Using equation (\ref{eq:9}),
with $\left[p_{i}^{nc},A_{i}^{\star}\right]=0$, equation (\ref{eq:8})
becomes
\begin{equation}
\left\{ \frac{\left(\mathbf{p}^{nc}\right)^{2}}{2m_{e}}-\frac{e\mathbf{p}^{nc}.\mathbf{A}^{\star}}{cm_{e}}+\frac{e^{2}\left(\mathbf{A}^{\star}\right)^{2}}{2c^{2}m_{e}}+\mu_{B}\sigma_{z}B\right\} \bar{\psi}=E\bar{\psi},\label{eq:10}
\end{equation}
where $\sigma_{z}=\pm1$. It is easy to check that  
\begin{equation}
\left(\mathbf{p}^{nc}\right)^{2}=p_{x}^{2}+p_{y}^{2}-\frac{\eta}{\hbar}L_{z}+\frac{\eta^{2}}{4\hbar^{2}}\left(x^{2}+y^{2}\right),\label{eq:11-a}
\end{equation}
\begin{equation}
\left(\mathbf{A}^{\star}\right)^{2}=\frac{B^{2}}{4}\left\{ x^{2}+y^{2}+\frac{\Theta^{2}}{4\hbar^{2}}\left(p_{x}^{2}+p_{y}^{2}\right)-\frac{\Theta}{\hbar}L_{z}\right\} ,\label{eq:11-b}
\end{equation}
{\small{}
\begin{equation}
\mathbf{p}^{nc}.\mathbf{A}^{\star}=\frac{-B}{2}\left\{ \frac{\Theta}{2\hbar}\left(p_{x}^{2}+p_{y}^{2}\right)+\frac{\eta}{2\hbar}\left(y^{2}+x^{2}\right)-\left(1+\frac{\Theta\eta}{4\hbar^{2}}\right)L_{z}\right\} ,\label{eq:11-c}
\end{equation}
}with 
\begin{equation}
L_{z}=\left(x_{i}\times p_{i}\right)_{z}=p_{y}x-p_{x}y.\label{eq:15-1}
\end{equation}

Using the three-equations (\ref{eq:11-a}-\ref{eq:11-c}) above, the
Pauli equation reads 
\begin{equation}
\left\{ \frac{\left(p_{x}^{2}+p_{y}^{2}\right)}{2\tilde{m}}-\tilde{\omega}L_{z}+\frac{\tilde{m}\tilde{\omega}^{2}}{2}\left(x^{2}+y^{2}\right)+\mu_{B}\sigma_{z}B\right\} \bar{\psi}=E\bar{\psi},\label{eq:14}
\end{equation}
with {\large{}
\begin{equation}
\begin{array}{c}
\tilde{m}=\frac{m_{e}}{\left(1+\frac{e\Theta B}{4c\hbar}\right)^{2}},\;\tilde{\omega}=\frac{eB\hbar+c\eta}{2c\hbar\tilde{m}\left(1+\frac{e\Theta B}{4c\hbar}\right)},\\
\frac{1}{2}\tilde{m}\tilde{\omega}^{2}=\frac{1}{2m_{e}}\left(\frac{e\eta B}{2c\hbar}+\frac{\eta^{2}}{4\hbar^{2}}+\frac{e^{2}B^{2}}{c^{2}4}\right).
\end{array}\label{eq:15}
\end{equation}
} 

We assume that $\tilde{\omega}$ is the deformed cyclotron frequency,
where in $\Theta\rightarrow0$, $\eta\rightarrow0$  limits, $\tilde{\omega}$
is reduced to $\frac{\omega_{c}}{2}=\frac{eB}{2cm_{e}}$. 

On the other hand, in case of an atomic Hydrogen, the electron is
bound to a proton by the Coulomb potential $A_{0}^{\star}$, which
is given by 
\begin{equation}
A_{0}^{\star}=\frac{e}{4\pi\epsilon_{0}}\frac{e}{\sqrt{x^{2}+y^{2}+\frac{\Theta^{2}}{4\hbar^{2}}\left(p_{x}^{2}+p_{y}^{2}\right)-\frac{\Theta}{\hbar}L_{z}}}.\label{eq:16-1}
\end{equation}

Our system looks like a two-dimensional harmonic oscillator with an
additional interaction $(-\tilde{\omega}L_{z}+\mu_{B}\sigma_{z}B)$.
This system corresponds to the Landau level problem, it corresponds
to the motion of a charged particle in the $xy$ plane and subjected
to a uniform magnetic field (in the symmetric gauge) oriented along
the axis (Oz), which means the particle is in interaction with its
orbital and spin angular momentum. The Hamiltonian from equation (\ref{eq:14})
can be written as
\begin{equation}
H_{Pauli}^{nc}=H_{nc}^{ho}-\tilde{\omega}L_{z}+\mu_{B}\sigma_{z}B.\label{eq:18-1}
\end{equation}

This problem will be solved simply by introducing operators of creation
and annihilation of harmonic oscillator, thus we define
\begin{equation}
\begin{array}{c}
a=\frac{1}{2}\sqrt{\frac{\tilde{\omega}}{\hbar}}\left(x-iy\right)+\frac{i}{2}\sqrt{\frac{1}{\hbar\tilde{\omega}}}\left(p_{x}-ip_{y}\right),\\
a=\frac{1}{2}\sqrt{\frac{\tilde{\omega}}{\hbar}}\left(x+iy\right)+\frac{i}{2}\sqrt{\frac{1}{\hbar\tilde{\omega}}}\left(p_{x}+ip_{y}\right),
\end{array}\label{eq:18-2}
\end{equation}
satisfying the following commutation relations
\begin{equation}
\left[a,a^{\dagger}\right]=\left[b,b^{\dagger}\right]=1.\label{eq:20-1}
\end{equation}

In terms of the ladder operators (\ref{eq:18-2}), our Hamiltonian
terms can be re-written as
\begin{equation}
L_{z}=\hbar\left(a^{\dagger}a-b^{\dagger}b\right),\label{eq:21-1}
\end{equation}
\begin{equation}
\begin{array}{ccc}
H_{nc}^{ho} & = & \hbar\tilde{\omega}\left(a^{\dagger}a+b^{\dagger}b+1\right)-\hbar\tilde{\omega}\left(a^{\dagger}a-b^{\dagger}b\right)\\
 & = & 2\hbar\tilde{\omega}\left(b^{\dagger}b+\frac{1}{2}\right).
\end{array}\label{eq:21-2}
\end{equation}

Eigenstates of our Hamiltonian are labeled by the number $j$ of excitation
quanta of the oscillator $a$, and the number $n$ of excitation quanta
of the oscillator $b$, 
\begin{equation}
a^{\dagger}a\mid n,j>=j\mid n,j>\text{ and }b^{\dagger}b\mid n,j>=n\mid n,j>,\label{eq:23-1}
\end{equation}
where both $n$ and $j$ can take on any positive integer value. Therefore,
our Pauli system becomes
\begin{equation}
\left\{ \hbar\tilde{\omega}\left(3b^{\dagger}b-a^{\dagger}a+1\right)+\mu_{B}\sigma_{z}B\right\} \mid n,j>=E\mid n,j>,\label{eq:23-2}
\end{equation}
with $\pm1$ are the eigenvalue of $\sigma_{z}$, therefore, our system
energy spectrum (discretely quantised) reads
\begin{equation}
E=\hbar\tilde{\omega}\left(3n-j+1\right)\pm\mu_{B}B.\label{eq:23-3}
\end{equation}

The effect of the phase-space noncommutativity is reduced in $\tilde{\omega}$.
Thus, by using equation (\ref{eq:15}) we have 
\begin{equation}
E_{n,j}\left(\Theta,\eta\right)=\frac{eB\hbar+c\eta}{2cm_{e}}\left(1+\frac{e\Theta B}{4c\hbar}\right)\left(3n-j+1\right)\pm\mu_{B}B.\label{eq:23-4}
\end{equation}

The above spectrum is a bit different from that obtained in ref. \cite{key-24}
in the limit of $\eta\rightarrow0$. However, the slight difference
is because the authors considered the magnetic field term proportional
to $\frac{1}{2m_{e}}$. In the limits of $\Theta\rightarrow0$ and
$\eta\rightarrow0$, the NC energy spectrum becomes commutative one,
i.e. commutative Landau system \cite{key-25}. 

After finding the energy spectrum, we now find the wave function.
The time-independent Pauli equation (\ref{eq:14}) reads
\begin{equation}
\left\{ \frac{-\hbar^{2}}{2\tilde{m}}\left(\frac{d^{2}}{dx^{2}}+\frac{d^{2}}{dy^{2}}\right)-\tilde{\omega}L_{z}+\frac{\tilde{\omega}^{2}\tilde{m}}{2}r^{2}+\mu_{B}\sigma_{z}B\right\} \bar{\psi}=E\bar{\psi},\label{eq:16}
\end{equation}
with $r=\sqrt{x^{2}+y^{2}}$. Now, we use cylindrical coordinates
$\left(r,\Phi\right)$ to solve the corresponding Pauli equation.
The wave function $\bar{\psi}\left(r,\Phi\right)$ can be given by
\begin{equation}
\bar{\psi}\left(r,\Phi\right)=R\left(r\right)e^{im\Phi},\label{eq:17}
\end{equation}
knowing that $m=0,\pm1,\pm2,\pm3,...$ are the eigenvalues of the
orbital angular momentum operator $L_{z}$. By replacing equation
(\ref{eq:17}) into equation (\ref{eq:16}), we obtain {\small{}
\begin{equation}
\left\{ \frac{-\hbar^{2}}{2\tilde{m}}\left(\frac{1}{r}\frac{d}{dr}+\frac{d^{2}}{dr}-\frac{m^{2}}{r^{2}}\right)-\tilde{\omega}m+\frac{\tilde{\omega}^{2}\tilde{m}}{2}r^{2}+\mu_{B}\sigma_{z}B\right\} R=ER.\label{eq:18}
\end{equation}
}{\small \par}

We can now solve the two-dimensional Pauli equation by assuming a
new functional form for $R\left(r\right)$ which is 
\begin{equation}
R\left(r\right)=\frac{\mathcal{Y}\left(r\right)}{\sqrt{r}},\label{eq:19}
\end{equation}
and choosing the lower eigenvalue of $\sigma_{z}$, thus the resulting
equation for $\mathcal{Y}\left(r\right)$ is
\begin{equation}
\left\{ -\frac{1}{2\tilde{m}}\frac{d^{2}}{dr^{2}}+\frac{m^{2}-\frac{1}{4}}{2\tilde{m}r^{2}}-m\tilde{\omega}+\frac{\tilde{m}\tilde{\omega}^{2}r^{2}}{2}-\mu_{B}B\right\} \mathcal{Y}=E\mathcal{Y}.\label{eq:20}
\end{equation}

We have used natural units with $\hbar=c=e=1$ to simplify more in
this part only. Let us find the corresponding wave functions. We can
rewrite the left-hand side of the equation (\ref{eq:20}) in the form
$\mathcal{A}^{\dagger}\mathcal{A}$, with
\begin{equation}
\begin{cases}
\mathcal{A}= & \frac{d}{dr}-\left(\frac{\left|m\right|+\frac{1}{2}}{r}\right)+\tilde{m}\tilde{\omega}r,\\
\mathcal{A}^{\dagger}= & -\frac{d}{dr}-\left(\frac{\left|m\right|-\frac{1}{2}}{r}\right)+\tilde{m}\tilde{\omega}r,
\end{cases}\label{eq:23}
\end{equation}
where the decomposition holds if $\left|m\right|\leq0$, which leads
to $E_{0}\geq0$. $E_{0}=0$ exists if and only if the solution of
the equation $\mathcal{A}\psi_{0}\left(r\right)=0$, thus
\begin{equation}
\left(\frac{d}{dr}-\left(\frac{\left|m\right|+\frac{1}{2}}{r}\right)+\tilde{m}\tilde{\omega}r\right)\bar{\psi}_{0}\left(r\right)=0,\label{eq:26}
\end{equation}
hence
\begin{equation}
\frac{d\bar{\psi}_{0}\left(r\right)}{\bar{\psi}_{0}\left(r\right)}=\left(\frac{\left|m\right|+\frac{1}{2}}{r}-\tilde{m}\tilde{\omega}r\right)dr.\label{eq:27}
\end{equation}

We solve the above equation to find 
\begin{equation}
\bar{\psi}_{0}\left(r\right)=k_{0}r^{\left|m\right|+\frac{1}{2}}\text{exp}\left[-\frac{1}{2}\tilde{m}\tilde{\omega}r^{2}\right],\label{eq:28}
\end{equation}
where $k_{0}$ is the normalization factor. In the limits of $\Theta\rightarrow0$
and $\eta\rightarrow0$, the above result reduces to the commutative
one, which corresponds to that of ref. \cite{key-26}, and it is given
by 
\begin{equation}
\bar{\psi}_{0}\left(r\right)=k_{0}r^{\left|m\right|+\frac{1}{2}}\text{exp}\left[-\frac{B}{2}r^{2}\right].\label{eq:29}
\end{equation}

\section{{\normalsize{}The semi-classical partition function and thermodynamic
properties in noncommutative phase-space}}

In the language of the classical treatment, we investigate the thermodynamic
properties of the two-dimensional noncommutative Pauli equation using
the semi-classical partition function. We initially focus on the calculation
of the semi-classical partition function $\mathcal{Z}$. Our studied
system is semi-classical where the Hamiltonian is split as follows
\begin{equation}
\mathcal{H}_{Pauli}^{2D}=H_{classic}+H_{ncl,\sigma},\label{eq:30-1}
\end{equation}
with $H_{ncl,\sigma}=\mu_{B}\sigma_{z}B$. Therefore, the noncommutative
partition function is separable into two independent parts as followed
recently in our work in Ref. \cite{key-27} 
\begin{equation}
\mathcal{Z}=Z_{Cl}Z_{ncl},\label{eq:30}
\end{equation}
where $Z_{ncl}$ is the non-classical part of the partition function.
To study our non-classical partition function, we assume that the
passage between noncommutative classical mechanics and noncommutative
quantum mechanics can be realized through the following generalized
Dirac quantization condition \cite{key-28,key-29}
\begin{equation}
\left\{ f,g\right\} =\frac{1}{i\hbar}\left[F,G\right],\label{eq:33-1}
\end{equation}
where $F$, $G$ stand for the operators associated with classical
observables $f$, $g$ and $\left\{ ,\right\} $ stands for Poisson
bracket. Using the condition above, we obtain from Eq.(\ref{eq:1-2})
\begin{equation}
\begin{array}{ccc}
\left\{ x_{j}^{nc},x_{k}^{nc}\right\}  & = & \Theta_{jk},\\
\left\{ p_{j}^{nc},p_{k}^{nc}\right\}  & = & \eta_{jk},\\
\left\{ x_{j}^{nc},p_{k}^{nc}\right\}  & = & \delta_{jk}+\frac{\Theta_{jl}\eta_{lk}}{4\hbar^{2}}=\delta_{jk}.
\end{array}\label{eq:33-2}
\end{equation}

It is important to mention that in terms of the classical limit, $\frac{\Theta\eta}{4\hbar^{2}}\ll1$
(check ref. \cite{key-29}), thus $\left\{ x_{j}^{nc},p_{k}^{nc}\right\} =\delta_{jk}$.
Now based on the proposal that noncommutative observables $F^{nc}$
corresponding to the commutative one $F(x,p)$ can be defined by \cite{key-23,key-30,key-31}
\begin{equation}
F^{nc}=F(x^{nc},p^{nc}),\label{eq:33-3}
\end{equation}
and for non-interacting particles, the classical partition function
in noncommutative phase-space for $N$ particles is written as follows
\cite{key-27,key-28} 
\begin{equation}
Z_{cl}=\frac{1}{N!\left(2\pi\tilde{\hbar}\right)^{2N}}\int e^{-\beta H_{classic}}d^{2N}x^{nc}d^{2N}p^{nc}.\label{eq:31}
\end{equation}

With $\frac{1}{N!}$ is Gibbs's correction factor, considered due
to accounting for indistinguishability, which means that there are
$N!$ ways of arranging $N$ particles at $N$ sites. $\frac{1}{\bar{\hbar}^{2}}$
is the appropriate factor that makes the volume of the noncommutative
phase-space dimensionless. $\beta$ defined as $\frac{1}{K_{B}T}$,
$K_{B}$ is the Boltzmann constant. 

Using equation (\ref{eq:30}) we may derive the important thermodynamic
quantities such as the Helmholtz free energy 
\begin{equation}
F=-\frac{1}{\beta}\text{ln}\mathcal{Z},\label{eq:35}
\end{equation}
and the average energy 
\begin{equation}
U\equiv N\left\langle \varepsilon\right\rangle =-\frac{\partial}{\partial\beta}\text{ln}\mathcal{Z},\label{eq:36}
\end{equation}
where $\varepsilon$ is the mean energy, which is given by $-\frac{\partial}{\partial\beta}\text{ln}\mathcal{Z}_{1}$.
Also the specific heat (heat capacity) 
\begin{equation}
C_{v}=\frac{\partial}{\partial T}\left\langle \varepsilon\right\rangle ,\label{eq:37}
\end{equation}
 as well the entropy 
\begin{equation}
S=-\frac{\partial F}{\partial T}=-\frac{K_{B}\text{ln}\mathcal{Z}}{\beta^{2}}+\frac{1}{\beta}\frac{\partial}{\partial T}\text{ln}\mathcal{Z}.\label{eq:38-1}
\end{equation}

Now for a single particle, the noncommutative classical partition
function is given by
\begin{equation}
Z_{cl,1}=\frac{1}{\tilde{\hbar}^{2}}\int e^{-\beta H_{classic}\left(x,p\right)}d^{2}x^{nc}d^{2}p^{nc},\label{eq:39}
\end{equation}
where $d^{2}$ is a shorthand notation serving as a reminder that
the $x$ and $p$ are vectors in two-dimensional phase-space. The
relation between equation (\ref{eq:31}) and (\ref{eq:39}) is given
by the following formula
\begin{equation}
Z_{cl}=\frac{\left(Z_{cl,1}\right)^{N}}{N!}.\label{eq:40}
\end{equation}

From equation (\ref{eq:4}), we simply have 
\begin{equation}
d^{2}x^{nc}d^{2}p^{nc}=\left(1-\frac{\Theta\eta}{4\hbar^{2}}\right)d^{2}xd^{2}p,\label{eq:42}
\end{equation}
and we have $\tilde{\hbar}\sim\triangle x^{nc}\triangle p^{nc}$,
which is given by 
\begin{equation}
\tilde{\hbar}^{2}=\hbar^{2}\left(1+\frac{\Theta\eta}{2\hbar^{2}}\right)+\mathcal{O}\left(\Theta^{2}\eta^{2}\right).\label{eq:43}
\end{equation}

Following now equation (\ref{eq:39}) we express the single particle
noncommutative classical partition function as 
\begin{equation}
Z_{cl,1}=\frac{1}{\tilde{\hbar}^{2}}\int e^{-\beta\left[\frac{p_{x}^{2}+p_{y}^{2}}{2\tilde{m}}-\tilde{\omega}L_{z}+\frac{\tilde{m}\tilde{\omega}^{2}}{2}\left(x^{2}+y^{2}\right)\right]}d^{2}x^{nc}d^{2}p^{nc}.\label{eq:50}
\end{equation}

We should mention again, as we emphasized in our previous work \cite{key-27}
that within the classical limit it is always possible to factorize
our Hamiltonian into momentum and position terms. Thus, we have 

\begin{equation}
Z_{cl,1}=\frac{1}{\tilde{\hbar}^{2}}\int e^{-\beta\frac{\left(p_{x}^{2}+p_{y}^{2}\right)}{2\tilde{m}}}e^{-\beta\frac{\tilde{m}\tilde{\omega}^{2}}{2}\left(x^{2}+y^{2}\right)}e^{\beta\tilde{\omega}L_{z}}d^{2}p^{nc}d^{2}x^{nc}.\label{eq:62}
\end{equation}

Using the same method used in our previous work \cite{key-27}, which
deponds on expanding exponentials containing $\tilde{\omega}$, and
by considering terms up to the second-order of $\tilde{\omega}$,
we find
\begin{equation}
\begin{array}{c}
Z_{cl,1}=\frac{1}{\tilde{\hbar}^{2}}\int e^{-\frac{\beta}{2}\left[\frac{p_{x}^{2}+p_{y}^{2}}{\tilde{m}}\right]}\left(1+\beta\tilde{\omega}L_{z}+\frac{1}{2}\beta^{2}\tilde{\omega}^{2}L_{z}^{2}\right)\\
\times\left(1-\beta\tilde{\omega}^{2}\frac{\tilde{m}}{2}\left(x^{2}+y^{2}\right)\right)d^{2}p^{nc}d^{2}x^{nc}.
\end{array}\label{eq:63}
\end{equation}

Knowing that
\begin{equation}
\left(1-\frac{\Theta\eta}{4\hbar^{2}}\right)\left(1-\frac{\Theta\eta}{2\hbar^{2}}\right)=1-\frac{3\Theta\eta}{4\hbar^{2}}+\mathcal{O}\left(\Theta^{2}+\eta^{2}\right),\label{eq:48-1}
\end{equation}
thus we have the convenient expression of $Z_{cl,1}${\large{}
\begin{equation}
\begin{array}{c}
Z_{cl,1}=\frac{1-\frac{3\Theta\eta}{4\hbar^{2}}}{h^{2}}\int e^{-\frac{\beta}{2}\left[\frac{p_{x}^{2}+p_{y}^{2}}{\tilde{m}}\right]}d^{2}pd^{2}x\\
+\frac{\left(1-\frac{3\Theta\eta}{4\hbar^{2}}\right)}{h^{2}}\beta\tilde{\omega}\int e^{-\frac{\beta}{2}\left[\frac{p_{x}^{2}+p_{y}^{2}}{\tilde{m}}\right]}L_{z}d^{2}pd^{2}x\\
+\frac{\left(1-\frac{3\Theta\eta}{4\hbar^{2}}\right)}{h^{2}}\beta^{2}\tilde{\omega}^{2}\int e^{-\frac{\beta}{2}\left[\frac{p_{x}^{2}+p_{y}^{2}}{\tilde{m}}\right]}L_{z}^{2}d^{2}pd^{2}x\\
-\frac{\left(1-\frac{3\Theta\eta}{4\hbar^{2}}\right)}{h^{2}}\beta\tilde{\omega}^{2}\int e^{-\frac{\beta}{2}\left[\frac{p_{x}^{2}+p_{y}^{2}}{\tilde{m}}\right]}\left(x^{2}+y^{2}\right)d^{2}pd^{2}x.
\end{array}\label{eq:64}
\end{equation}
}{\large \par}

In the right-hand side of the above equation, the second integral
goes to zero, the third and fourth integrals cancel each other, then
by using the known integral of Gaussian function $\int e^{-ax^{2}}dx=\sqrt{\frac{\pi}{a}}$,
we find 
\begin{equation}
Z_{cl,1}=\frac{1-\frac{3\Theta\eta}{4\hbar^{2}}}{h^{2}}\int d^{2}xe^{-\frac{\beta}{2}\left[\frac{p_{x}^{2}+p_{y}^{2}}{\tilde{m}}\right]}d^{2}p=\frac{l^{2}\left(1-\frac{3\Theta\eta}{4\hbar^{2}}\right)}{\varLambda^{2}\left(1+\frac{e\Theta B}{4c\hbar}\right)^{2}},\label{eq:65}
\end{equation}
 with $\int d^{2}x=l^{2}$, $\varLambda=h\left(2\pi m_{e}K_{B}T\right)^{-\frac{1}{2}}$
are the area and the thermal de Broglie wavelength respectively. 

We also propose another method based on the substitution of variables
with the the Jacobian matrix to compute the integral (\ref{eq:50}),
explained in \textbf{Appendix \ref{A}, }which gives the same results.

The non-classical partition function for a $N$ particle is given
by
\begin{equation}
Z_{ncl}=Z_{ncl,1}^{N}=\left(\sum_{\sigma_{z}=\pm1}e^{\beta\mu_{B}\sigma_{z}B}\right)^{N}=2^{N}\text{cosh}{}^{N}\left(\beta\mu_{B}B\right).\label{eq:67}
\end{equation}

An important point to note is for a canonical ensemble that is classical
and discrete, the canonical partition function is defined using sum
as in the case of $H_{ncl,\sigma}$. But for a canonical ensemble
that is classical and continuous, the canonical partition function
is defined using integral.

Finally, the Pauli partition function (\ref{eq:30}) for a system
of $N$ particles in a two-dimensional noncommutative phase-space
is 
\begin{equation}
\mathcal{Z}=\frac{2^{N}l^{2N}}{\varLambda^{2N}N!}\frac{\left(1-\frac{3\Theta\eta}{4\hbar^{2}}\right)^{N}}{\left(1+\frac{eB\Theta}{8c\hbar}\right)^{2N}}\text{cosh}{}^{N}\left(\beta\mu_{B}B\right).\label{eq:50-2}
\end{equation}

In the vanishing limit of the noncommutativity, i.e. $\Theta\rightarrow0$,
$\eta\rightarrow0$, the expression of $\mathcal{Z}$ reduces to that
of the usual commutative phase-space, which is 
\begin{equation}
\mathcal{Z}=\frac{2^{N}l^{2N}}{\varLambda^{2N}N!}\text{cosh}{}^{N}\left(\beta\mu_{B}B\right).\label{eq:51}
\end{equation}

Following the relations (\ref{eq:35}, \ref{eq:36}, \ref{eq:37},
\ref{eq:38-1}) and as a consequence of equation (\ref{eq:50-2}),we
express the thermodynamic quantities in noncommutative phase-space,
thus we have
\begin{equation}
F^{nc}=-\frac{N}{\beta}\text{ln}\left[\frac{2l^{2}}{\varLambda^{2}}\frac{\left(1-\frac{3\Theta\eta}{4\hbar^{2}}\right)}{\left(1+\frac{eB\Theta}{8c\hbar}\right)^{2}}\text{cosh}\left(\beta\mu_{B}B\right)\right]+\frac{1}{\beta}\text{ln}N!,\label{eq:52}
\end{equation}
where $\text{ln}N!\approx N\text{ln}N-N$. 
\begin{equation}
\begin{array}{c}
S^{nc}=\frac{K_{B}N}{\beta^{2}}\left\{ 1+\frac{\text{ln}N!}{N}+\beta\mu_{B}B\text{tanh}\left(\beta\mu_{B}B\right)\right.\\
\left.-\text{ln}\left[\frac{2l^{2}}{\varLambda^{2}}\frac{\left(1-\frac{3\Theta\eta}{4\hbar^{2}}\right)}{\left(1+\frac{eB\Theta}{8c\hbar}\right)^{2}}\text{cosh}\left(\beta\mu_{B}B\right)\right]\right\} ,
\end{array}\label{eq:53}
\end{equation}
\begin{equation}
U^{nc}=N\left[\frac{1}{\beta}-\mu_{B}B\text{tanh}\left(\beta\mu_{B}B\right)\right],\label{eq:55}
\end{equation}
\begin{equation}
\left\langle \varepsilon^{nc}\right\rangle =\frac{1}{\beta}-\mu_{B}B\text{tanh}\left(\beta\mu_{B}B\right),\label{eq:55-1}
\end{equation}
\begin{equation}
C_{v}^{nc}=-K_{B}\left[\frac{1}{\beta^{2}}+\frac{\left(\mu_{B}B\right)^{2}}{\text{cosh}^{2}\left(\beta\mu_{B}B\right)}\right].\label{eq:56}
\end{equation}

In the vanishing limit of the noncommutativity, the result of this
paper will be reduced to that of commutative phase space. Such as
\begin{equation}
F=-\frac{N}{\beta}\text{ln}\left[\frac{2l^{2}}{\varLambda^{2}}\text{cosh}\left(\beta\mu_{B}B\right)\right]+\frac{1}{\beta}\text{ln}N!,\label{eq:56-1}
\end{equation}
as well
\begin{equation}
\begin{array}{c}
S=\frac{K_{B}N}{\beta^{2}}\left\{ 1+\frac{\text{ln}N!}{N}+\beta\mu_{B}B\text{tanh}\left(\beta\mu_{B}B\right)\right.\\
\left.-\text{ln}\left[\frac{2l^{2}}{\varLambda^{2}}\text{cosh}\left(\beta\mu_{B}B\right)\right]\right\} .
\end{array}\label{eq:57}
\end{equation}

Through further derivatives, we can go deeper and calculate the rest
of the thermodynamic properties using the obtained partition function.
Such as temperature $T$, pressure $P$, the magnetization $\left\langle M\right\rangle $
and chemical potential $\mu$.

\section{{\normalsize{}Conclusion}}

In this work, we have discussed the problem of a charged particle
with a spin in interaction with an electromagnetic field moving in
a two-dimensional noncommutative phase-space, by considering a constant
magnetic field perpendicular to the plane. The approach that we have
took to map the noncommutative problem to the equivalent commutative
one is the Bopp-shift transformation. We found the energy spectrum,
which is discretely quantised and the wave function of the two-dimensional
noncommutative Pauli equation. The effect of the noncommutative parameters
on the energy spectrum and wave function is significant. In addition,
according to equation (\ref{eq:15}), we can see an emerge of a modified
frequency $\tilde{\omega}$, which represents the effect of the noncommutativity
on the cyclotron frequency. Knowing that in the limits of $\Theta\text{\textrightarrow}0$
and $\eta\text{\textrightarrow}0$, the noncommutative results reduce
to that of usual commutative phase-space.

Furthermore, by using the classical treatment, some classical statistical
quantities are determined in the two-dimensional noncommutative phase-space
using a semi-classical partition function from the Pauli system of
the one-particle and N-particle systems in two dimensions, all according
to the canonical ensemble theory. It is shown that Helmholtz free
energy and entropy were significantly affected by the noncommutativity
of the phase space. In contrast, specific heat and average energy
showed non-dependence on noncommutativity.

Note that the result (\ref{eq:65}) is of the classical Maxwell-Boltzmann
gas, as this happens upon the classical calculation of Landau problem.
On the other hand, the quantum partition function for the Landau problem
represents de Haas-van Alphen effect. 

The results of the present work can be used to expand the study on
a possible generalization to make the consideration of \textbf{anyons},
i.e., particles with arbitrary non-integer spin, which can exist in
two-dimensional space.

\begin{acknowledgments}
The author is very thankful to the anonymous referee for pointing
out mistakes and suggesting improvements.
\end{acknowledgments}

\appendix

\section{\label{A}Integration using the substitution of multiple variables
with Jacobian matrix}

Here is a method based on the substitution of multiple variables with
the determinant of the Jacobian matrix to compute the integral (\ref{eq:50}).

The substitution of multiple variables is as follows
\begin{equation}
\begin{cases}
x=x,\\
y=y,\\
P_{x}=p_{x}+\tilde{m}\tilde{\omega}y,\\
P_{y}=p_{y}-\tilde{m}\tilde{\omega}x,
\end{cases}\label{eq:a-1}
\end{equation}
where the integral (\ref{eq:50}) is
\begin{equation}
\begin{array}{ccc}
\text{Int} & = & \frac{1}{\tilde{\hbar}^{2}}\int e^{-\beta\left[\frac{p_{x}^{2}+p_{y}^{2}}{2\tilde{m}}-\tilde{\omega}L_{z}+\frac{\tilde{m}\tilde{\omega}^{2}}{2}\left(x^{2}+y^{2}\right)\right]}d^{2}x^{nc}d^{2}p^{nc}\\
 & = & \frac{1-\frac{3\Theta\eta}{4\hbar^{2}}}{h^{2}}\int e^{-\beta\left[\frac{p_{x}^{2}+p_{y}^{2}}{2\tilde{m}}-\tilde{\omega}L_{z}+\frac{\tilde{m}\tilde{\omega}^{2}}{2}\left(x^{2}+y^{2}\right)\right]}dxdydp_{x}dp_{y}\\
 & = & \frac{1-\frac{3\Theta\eta}{4\hbar^{2}}}{h^{2}}\int e^{-\frac{\beta}{2\tilde{m}}\left[\left(P_{x}^{2}+P_{y}^{2}\right)\right]}\left|\text{Det}\text{J}\left(x,y,P_{x},P_{y}\right)\right|dxdydP_{x}dP_{y}.
\end{array}\label{eq:a-2}
\end{equation}

The corresponding Jacobian matrix is 
\begin{equation}
\text{J}\left(x,y,P_{x},P_{y}\right)=\begin{bmatrix}\frac{\partial x}{\partial x} & \frac{\partial x}{\partial y} & \frac{\partial x}{\partial p_{x}} & \frac{\partial x}{\partial p_{y}}\\
\frac{\partial y}{\partial x} & \frac{\partial y}{\partial y} & \frac{\partial y}{\partial p_{x}} & \frac{\partial y}{\partial p_{y}}\\
\frac{\partial P_{x}}{\partial x} & \frac{\partial P_{x}}{\partial y} & \frac{\partial P_{x}}{\partial p_{x}} & \frac{\partial P_{x}}{\partial p_{y}}\\
\frac{\partial P_{y}}{\partial x} & \frac{\partial P_{y}}{\partial y} & \frac{\partial P_{y}}{\partial p_{x}} & \frac{\partial P_{y}}{\partial p_{y}}
\end{bmatrix}=\begin{bmatrix}1 & 0 & 0 & 0\\
0 & 1 & 0 & 0\\
0 & \tilde{m}\tilde{\omega} & 1 & 0\\
-\tilde{m}\tilde{\omega} & 0 & 0 & 1
\end{bmatrix}.\label{eq:a-3}
\end{equation}

The determinant of the Jacobian matrix J is 
\begin{equation}
\text{Det}\text{J}\left(x,y,P_{x},P_{y}\right)=\begin{vmatrix}1 & 0 & 0 & 0\\
0 & 1 & 0 & 0\\
0 & \tilde{m}\tilde{\omega} & 1 & 0\\
-\tilde{m}\tilde{\omega} & 0 & 0 & 1
\end{vmatrix}=1\begin{vmatrix}1 & 0 & 0\\
\tilde{m}\tilde{\omega} & 1 & 0\\
0 & 0 & 1
\end{vmatrix}=\begin{vmatrix}1 & 0\\
0 & 1
\end{vmatrix}=1.\label{eq:a-4}
\end{equation}

Therefore, the integral (\ref{eq:a-1}) becomes

\begin{equation}
\begin{array}{ccc}
 & = & \frac{1-\frac{3\Theta\eta}{4\hbar^{2}}}{h^{2}}\int e^{-\frac{\beta}{2\tilde{m}}\left[\left(P_{x}^{2}+P_{Y}^{2}\right)\right]}dP_{x}dP_{y}\int dxdy\\
 & = & \frac{1-\frac{3\Theta\eta}{4\hbar^{2}}}{h^{2}\left(1+\frac{e\Theta B}{4c\hbar}\right)^{2}}\frac{2\pi m_{e}}{\beta}\int dxdy,
\end{array}\label{eq:a-5}
\end{equation}
then by using the known integral of Gaussian function $\int e^{-a\left(x^{2}+y^{2}\right)}dx=\frac{\pi}{a}$,
with $\int d^{2}x=l^{2}$, $\varLambda=h\left(2\pi m_{e}K_{B}T\right)^{-\frac{1}{2}}$
we find 
\begin{equation}
\text{Int}=\frac{l^{2}\left(1-\frac{3\Theta\eta}{4\hbar^{2}}\right)}{\varLambda^{2}\left(1+\frac{e\Theta B}{4c\hbar}\right)^{2}}.\label{eq:a-6}
\end{equation}

\end{document}